\documentclass[aps,superscriptaddress,twocolumn]{revtex4}
\usepackage{graphicx}
\usepackage{dcolumn}
\usepackage{bm}
\usepackage{colordvi}
\usepackage{color}

\begin{document}

\title{Electric dipole spin resonance at shallow donors in quantum
wires}
\author{D.V. Khomitsky}
\email{khomitsky@phys.unn.ru}
\affiliation{Department of Physics, National Research Lobachevsky State University of
Nizhni Novgorod, 603950 Gagarin Avenue 23, Nizhny Novgorod, Russian
Federation}
\author{E.A. Lavrukhina}
\affiliation{Department of Physics, National Research Lobachevsky State University of
Nizhni Novgorod, 603950 Gagarin Avenue 23, Nizhny Novgorod, Russian
Federation}
\author{E.Ya. Sherman}
\affiliation{Department of Physical Chemistry, The University of the Basque Country,
48080 Bilbao, Spain}
\affiliation{IKERBASQUE Basque Foundation for Science, Bilbao, Spain}

\begin{abstract}
Electric dipole spin resonance is studied  theoretically for a shallow donor formed in a
nanowire with spin-orbit coupling in a magnetic field. Such system
may represent a donor-based qubit. The single discrete energy level of the donor is 
accompanied by the set of continuum states, which provide a
non-trivial interplay for the picture of electric dipole spin resonance
driven by an external monochromatic field. Nonlinear dependencies of spin flip
time as well as of the coordinate mean values on the electric field
amplitude are observed, demonstrating the significance of coupling to the
continuum for spin-based qubits manipulation in nanostructures.
\end{abstract}

\date{\today}
\maketitle

\section{Introduction}

Electric dipole spin resonance (EDSR), that is the ability to manipulate
spins of charge carriers by electic rather than by a magnetic field, is
one of the most distinctive features of spin-orbit coupling (SOC). Being
theoretically predicted \cite{Rashba1960} and initially experimentally observed for itinerant 
electrons in bulk crystals \cite{Bemski1960,Bell1962,Bratashevskii1963}, soon it was studied theoretically in detail for electrons
localized on donors \cite{Rashba1964a} and for holes on the acceptor centers \cite{Rashba1964b}. 
More recently, it was shown that the EDSR is a powerful tool for 
spin manipulation in quantum wells \cite{Rashba2003} and other two-dimensional heterostructures with spin-orbit 
coupling \cite{Azpiroz}. In addition, the EDSR can be used to manipulate electic current 
in low-dimensional conductors \cite{Sadreev}. 
Observation of the EDSR in quantum dots \cite{Nowack2007} opened a venue for their applications in
spin-based quantum computing, where spin of a carrier localized in a quantum dot is considered as a
qubit. Single electron spin manipulation using hybrid semiconductor-superconductor systems 
has been proposed in Ref. [\onlinecite{Hu2012}].
As result, theoretical studies of the EDSR in quantum dots became the 
field attracting a great interest of researchers \cite{Golovach2006,Jiang2006,Borhani2012}. 
The importance of several effects such as nonlinear dynamics has been recognized and 
investigated \cite{Nowak2012,Romhanyi2015}. Spin manipulation by pulsed rather than periodic electric
fields has been studied, e.g. in Ref. [\onlinecite{Ban2012}] for designed pulses, and in Ref. [\onlinecite{Veszeli2018}]
for subcycle ones. Also, the EDSR can occur in quantum wires
\cite{Li2013} attracting a great deal of attention \cite{Nadj2010} in the information processing 
technologies. Another interesting example of the EDSR is presented by carbon nanotubes \cite{Osika2015}. 

Recently, other kinds of solid-state based qubits have been put forward. These states are related
to the Majorana fermions \cite{Oreg2010,Das2012,Brouwer2011,DeGottardi2013,Adagideli2014} 
in InSb-based nanowires, demonstrating rich disorder effects on the electron states \cite{Pitanti2011}. 
Another option is related to using shallow electron states as qubits \cite{Linpeng2016}. Here the studies of
the EDSR face a challenge, which has not been yet addressed in the
literature, since in the shallow donor systems the applied electric field can
couple the localized and delocalized states of the electron. Also, such a
modification of the states could be produced by the spin-orbit coupling. As a result,
the electron wavefunctions, spin-flip transition matrix elements, and the
entire spin dynamics become strongly modified. Shallow states in nanowires can be formed by 
charged donors screened by itinerant electrons in the surrounding metallic gates or in the two-dimensional 
electron gas used as a building element of the template structure. 

Note that the above listed approaches  considered either itinerant or localized states without taking into
account possible transitions to the continuum states, as can occur for the shallow donors. 
Here we address these issues and show how the entire picture of the EDSR for the shallow states 
is modified in the presence of the continuum.

This paper is organized as follows. In Section II we introduce the model of single-electron states
formed in a nanowire in the presence of a shallow donor potential, the SOC, and the magnetic field.
Both localized and delocalized states are considered, and the periodic potential of the driving electric 
field is introduced.
In Section III we present the results of the perturbation theory approach which allows to 
highlight several key features of the quantum states
including the ones stemming from the interplay of the SOC and the magnetic field.
In Section IV the numerical solution for the time-dependent problem with the driving is discussed, 
and the main results of the paper are presented for various
parameters of the model. In Section V we give our conclusions.

\section{Model Hamiltonian and external driving}

\begin{figure}[tbp]
\centering
\includegraphics*[width=65mm]{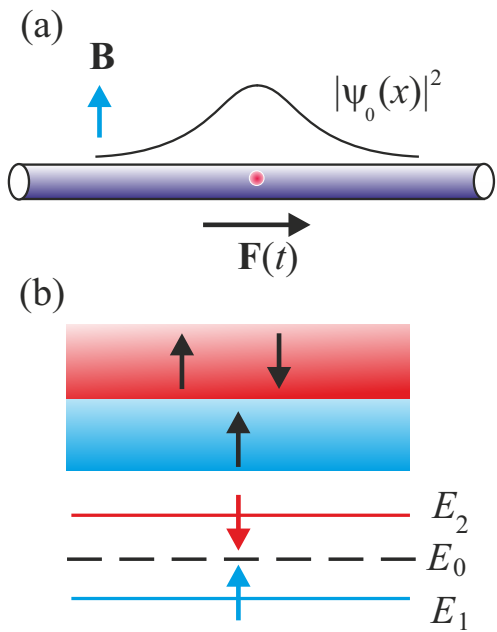}
\caption{ (a) A schematic layout of a shallow donor state formed in a quantum
wire. The vectors $\mathbf{B}$ and ${\mathbf{F}}(t)$ show the directions of
applied constant magnetic field and periodic electric field, respectively.
(b) Scheme of the Zeeman doublet $E_1$, $E_2$ and the continuum states shown in the shaded bands.}
\label{figsqw}
\end{figure}

We consider a narrow nanowire, elongated along the $x-$ axis, as shown in Fig. \ref{figsqw}(a),
where full three-dimensional (3D) electron wavefunction $\psi_{\rm 3D}(\mathbf{r})$ can be presented as a product
$\psi_{\rm 3D}(\mathbf{r})=\psi_{\rm 2D}(\mathbf{r}_{\perp})\Psi(x,t).$ 
Here $\psi_{\rm 2D}(\mathbf{r}_{\perp})$ corresponds to the 
ground state of the transverse motion with $\mathbf{r}_{\perp}=(y,z),$ and we will be interested only in 
the motion along the wire described by $\Psi(x,t).$
 
To characterize this one-dimensional motion, we begin with the 
following Hamiltonian using the effective mass approximation:
\begin{equation}
H_0=\frac{p^2}{2m}-\frac{U_0}{\cosh^2 (x/d)},  \label{h0}
\end{equation}
where the confinement by the donor is described as a potential with the 
effective width $d$ and the maximum 
depth $U_{0}$, $m$ is the electron effective mass, 
and $p=-i\hbar\partial/\partial x\equiv \hbar k$ is the momentum operator. 
We are interested here in shallow potentials satisfying condition of weak
binding such as $\hbar ^{2}/md^{2}\gg U_{0}.$ In general, in such potentials
the binding energy $E_{0}$ is determined by $E_{0}\sim -m(U_{0}d)^{2}/\hbar ^{2}.$
Omitting for the moment the spinor structure, we present the ground state wavefunction at distances
$|x|\gg d$ in the form \cite{Landau}:
\begin{equation}
\psi _{0}(x)=\frac{1}{\sqrt{l}}\exp\left(-\left\vert x\right\vert /l\right), \label{psi0}
\end{equation}
with the localization length $l=\sqrt{{\hbar^{2}}/{2m\left\vert E_{0}\right\vert }}\gg d.$ 

Next, we add the Rashba SOC $H_{R}=\alpha \sigma _{y}k$ and the constant
magnetic field $\mathbf{B}=(0,0,B)$ (see Fig. \ref{figsqw}) creating the Zeeman term in the
Hamiltonian, which now takes the following form:
\begin{equation}
H_{1}=H_{0}+\frac{\Delta}{2}\sigma_{z}+\alpha \sigma _{y}k.
\label{h1}
\end{equation}
Here the Zeeman splitting $\Delta=\mu_{B}gB,$ where $\mu_{B}$ is the Bohr magneton, and $g$ is the electron g-factor. 
Since we consider a narrow single quantum wire in the ground state transverse mode, we employ 
the basic form of the Rashba coupling. Description of the Rashba coupling in more complex nanowire systems 
can be found, e.g., in Refs. [\onlinecite{Mireles2001,Gelabert2010,Zainuddin2011,Xu2014}].

As a result, the ground state of (\ref{h0}) transforms into a Zeeman-split doublet containing the
only two localized eigenstates of (\ref{h1}), with spacing $E_{2}-E_{1}$ (see Fig. \ref{figsqw}(b)).
For $\alpha=0$ and $g<0$ the ground state $E_{1}$ is the spin-up state where the spin is 
parallel to the magnetic field, and the state $E_{2}$ is the spin-down state. These Zeeman partners may effectively
participate in the spin resonance driven by an external
periodic electric field applied along the wire, with the 
frequency $\omega=(E_{2}-E_{1})/\hbar$ and amplitude $F_{0}$, and described as a potential,
\begin{equation}
V(x,t)=eF_{0}x\sin \omega t, 
\label{Vxt}
\end{equation}%
where $e$ is the fundamental charge, being added to the Hamiltonian $H_{1}.$ 
The reason for this participation
is that  due to the presence of SOC
the eigenstates of (\ref{h1}) contain both spinor components.
Therefore, two states with different
signs of their $\sigma _{z}$ can still be coupled by the electric field
described by the position operator, creating the possibility of the
electric dipole spin resonance. 

As for the continuum states, they also split into Zeeman doublets as shown in Fig. \ref{figsqw}
by arrows in the bands above $E_1$ and $E_2$. At $\alpha=0$ and $g<0,$ the low-energy part of the continuum has spin up 
and is non-degenerate. The higher states are twofold degenerate where the spin-up 
state from the lower Zeeman energy is accompanied by the spin-down state from the higher Zeeman energy described 
by the eigenvalue set of continuum for $H_0$.

In the absence of the spin-orbit coupling, 
with the increase in the magnetic field, the discrete state $E_{2}$ reaches the bottom of
the continuum at the merging field $B_{\rm m}$ satisfying the condition
\begin{equation}
\mu_{B}|g|B_{\rm m}=|E_{0}|,  \label{bm}
\end{equation}%
and transition frequency $\omega=|E_{0}|\times(B/B_{\rm m})/\hbar.$
Below we will consider the fields lower than $B_{\rm m}$ in order to avoid the
direct overlap of the discrete and continuum states in the stationary
Hamiltonian (\ref{h1}). However, we will see that discrete and continuum
states interact dynamically during the evolution driven by periodic electric
field, as it will be discussed in the next Sections. 

\section{Spin-orbit coupling as a perturbation}

For a qualitative understanding of the electric dipole spin resonance here
we will apply the perturbation theory to highlight the effects of SOC and
the presence of continuum on the wavefunctions belonging only to
the lowest Zeeman-split doublet. We begin by considering the 
SOC term $\alpha \sigma _{y}k$ in
the Hamiltonian (\ref{h1}) as a perturbation, similarly to the approaches of Refs. [\onlinecite{Rashba1964a,Khaetsky}]. 
At $\alpha=0,$ we label the Zeeman-split doublet state as spin-up $(\lambda=1)$
or spin-down  $(\lambda=-1)$ state, respectively. 

The states of the discrete spectrum have the form:
\begin{equation}
\psi _{0}^{\lambda=-1} =\left[ 
\begin{array}{c}
0 \\ 
\psi _{0}(x)%
\end{array}%
\right] ,\qquad \psi _{0}^{\lambda=1}=\left[ 
\begin{array}{c}
\psi _{0}(x) \\ 
0%
\end{array}%
\right].  \label{psidisc0}
\end{equation}
The energies of this Zeeman doublet are $E_{0}^{[\lambda]}=E_{0}+\lambda{\Delta}/2,$
and the only shallow localized ground state is $\psi_{0}(x).$ At $\alpha=0$, 
$E_{0}^{\lambda=1} = E_{1}$ and $E_{0}^{\lambda=-1} = E_{2}.$  

The continuum states are modeled by the approximation of extremely far 
hard walls located at $|x|=L$ with $L\gg l,$ corresponding to the wire length $2L,$ 
which leads to the set 
of very densely located levels representing the continuum with required accuracy \cite{Lparameter}.
The spatially odd states
representing the continuous spectrum and coupled by the $k$ term to the  
even localized states are labeled by the discrete index $n$
in our model: 
\begin{equation}
\varphi _{n}^{\lambda=-1}=\left[ 
\begin{array}{c}
0 \\ 
\varphi _{n}(x)%
\end{array}%
\right] ,\qquad \varphi _{n}^{\lambda=1}=\left[ 
\begin{array}{c}
\varphi _{n}(x) \\ 
0%
\end{array}%
\right] ,  \label{psicont0}
\end{equation}%
with the above defined energies $E_{n}^{[\lambda]}=E_{n}+\lambda{\Delta }/{2}$ and 
\begin{equation}
\varphi _{n}(x)=\frac{1}{\sqrt{L}}\sin \left( k_{n}x\right) ,
\label{psicontappr}
\end{equation}%
where $E_{n}=\hbar ^{2}k_{n}^{2}/2m$ with $k_{n}=\pi n/L.$ 

When the SOC is turned on, the localized states (\ref%
{psidisc0}) forming the ground Zeeman doublet acquire the admixture from the
continuum states. In the leading order of the perturbation theory one can
present them as:
\begin{equation}
\psi ^{\lambda=-1} =\left[ 
\begin{array}{c}
\sum\limits_{n}a_{n}^{\lambda=-1}\varphi _{n}(x) \\ 
\psi _{0}(x)%
\end{array}%
\right], 
\psi ^{\lambda=1} =\left[ 
\begin{array}{c}
\psi _{0}(x) \\ 
\sum\limits_{n}a_{n}^{\lambda=1}\varphi _{n}(x)%
\end{array}%
\right]  \label{psifinal}
\end{equation}
where
\begin{equation}
a_{n}^{[\lambda]} =-\lambda\frac{\alpha}{E_{0}^{[\lambda]}-E_{n}^{[\lambda^{\prime}]}}
\int_{-\infty}^{\infty}\varphi _{n}(x) \psi_{0}^{\prime}(x)dx,  
\label{anstart}
\end{equation}
and for the wavefunction in Eq. (\ref{psi0}) one obtains
\begin{equation}
\int_{-\infty}^{\infty }\varphi _{n}(x)\psi_{0}^{\prime}(x)dx=
-\frac{2}{\sqrt{lL}}\frac{k_{n}l}{1+k_{n}^{2}l^{2}}.
\label{refint}
\end{equation}

By analyzing (\ref{anstart}) -(\ref{refint}) 
one can see that the coefficients $a_{n}^{({\pm }%
)}$ describing the admixture of continuum states for the localized
Zeeman-split doublet (\ref{psifinal}) are proportional to $\alpha.$ This is a simple confirmation of the role
played by the continuum states when the SOC is significant. The resulting spin-projected 
probability to find the electron in the continuum, $w_{\rm c}=\sum_{n}|a_{n}^{[\lambda]}|^{2},$ 
at $B=0$ is given by $w_{\rm c}=m(\alpha/\hbar)^{2}/|E_{0}|.$

For the following studies of the electric dipole spin resonance we will need the
matrix element of coordinate calculated between the states of our interest, that
is between $\psi^{\lambda=1}$ and $\psi^{\lambda=-1}$ in Eq. (\ref{psifinal}).
For the localized states (\ref{psifinal}) with opposite spins one obtains:
\begin{equation}
x_{\lambda\lambda^{\prime}} =  
\sum_{n}\left(a_{n}^{\lambda=-1}+a_{n}^{\lambda=1}\right)\int_{-\infty}^{\infty} \varphi _{n}(x)x\psi _{0}(x)dx.  
\end{equation}

Taking into account that
\begin{equation}
\int_{-\infty}^{\infty} \varphi _{n}(x)x\psi _{0}(x)dx=
4{\sqrt{\frac{l}{L}}}l\frac{k_{n}l}{(1+k_{n}^{2}l^{2})^{2}},
\label{xintegral}
\end{equation}
performing summation over $k_{n}$ and by using {(\ref{psi0}), (\ref{psicontappr}), and (\ref{anstart}), (\ref{refint})},
we arrive at
\begin{equation}
x_{\lambda\lambda^{\prime}} =  
\frac{\alpha }{\left\vert E_{0}\right\vert }\frac{1}{\xi ^{3}}%
\left[ \xi ^{2}+4\left( \sqrt{1-\xi }+\sqrt{1+\xi }-2\right) \right],  
\label{xnmfinal}
\end{equation}
with notation $\xi \equiv \Delta /\left\vert E_{0}\right\vert=B/B_{\rm m}.$ 
We consider $|\xi|<1$ to avoid direct overlap of the continuum
and localized states, where the energy levels acquire more complicated contributions. 
By expanding Eq. (\ref{xnmfinal}) by $\xi\ll 1$
we obtain $x_{\lambda\lambda^{\prime}}=-5\alpha\xi/16|E_{0}|.$
Note that the matrix element in Eq. (\ref{xnmfinal})
can be estimated in terms of two characteristic lengths of the model 
as $l\times l/l_{\rm so},$ with $l_{\rm so}={\hbar^{2}}/{m\alpha}$ being the spin precession length. 
While this is the result of the first order
perturbation theory which may no longer be applicable for large SOC, it
clearly shows the importance of considering the continuum states for the
system with shallow donor and strong SOC. Thus, the Zeeman-split
discrete states are SOC-coupled via the continuum. This coupling is
critical for the driven dynamics which will be analyzed in the next Section.

In a similar way we calculate the energy shift of the state of interest as
\begin{equation}
\Delta E_{\lambda} = -2m\frac{\alpha^{2}}{\hbar^{2}\xi^{2}}\left(\sqrt{1-\lambda\xi}-1\right)^{2}.  
\label{enshift}
\end{equation}
In the zero-field limit $\xi\rightarrow 0$ both $\Delta E^{\lambda=-1}$ and $\Delta E^{\lambda=1}$ 
behave as $-m\left(\alpha/\hbar\right)^{2}/2$
and their difference $\Delta E^{\lambda=-1}-\Delta E^{\lambda=1}\approx m\left(\alpha/\hbar\right)^{2}\xi/2.$
These corrections can be considered as a renormalization of the $g-$factor by
the spin-orbit coupling due to the presence of the continuum states and show that the perturbation theory 
is applicable at $m\left(\alpha/\hbar\right)^{2}/|E_{0}|\ll 1,$ that is at $l\ll l_{\rm so}$ and 
$|x_{\lambda\lambda^{\prime}}| \ll l.$

\section{Spin dynamics for periodic driving}

\subsection{Numerical basis states and model of the dynamics}

The numerically precise basis states $\phi _{n}$ and the energies $E_{n}^{(0)}$ of the
Hamiltonian (\ref{h0}) are found by the discretization 
on the $x-$ axis.  An eigenstate $\psi
_{l}(x)$ for (\ref{h1}) with the eigenenergy $E_{l}$ is constructed as a superposition of the basis
states $\phi _{n}$ of the Hamiltonian (\ref{h0}) with the coefficients
forming the two-component spinors:
\begin{equation}
\psi_l(x)= \sum_n \left[ 
\begin{array}{c}
a^l_n \\ 
b^l_n%
\end{array}
\right]\phi_n(x).  \label{psi}
\end{equation}
The spinor coefficients $a_{n}^{l}$ and $b_{n}^{l}$ are found from the
numerical diagonalization of (\ref{h1}) with $l=1,\dots,l_{\rm max},$ where
$l_{\max}$ is the basis size. 

Then, we solve numerically the nonstationary Schr\"{o}dinger equation
\begin{equation}
i\hbar\frac{\partial}{\partial t}\Psi(x,t)=H(x,t)\Psi(x,t), 
\label{scheq}
\end{equation}
with the Hamiltonian $H(x,t)=H_1+V(x,t)$, for the time-dependent wavefunction
\begin{equation}
\Psi(x,t)={\bm C}^{\rm T}(t)\bm{\psi}(x).  
\label{psit}
\end{equation}
In Eq. (\ref{psit}) ${\bm C}(t)$ is the vector with $l_{\rm max}$ components 
determined from (\ref{scheq}), and 
$\bm{\psi}(x)=\left(\psi_{1}(x),\dots,\psi_{l_{\rm max}}(x)\right)^{\rm T}$.
The Hamiltonian in (\ref{scheq}) is time-periodic, i.e. $H(x,t)=H(x,t+NT)$ for any integer $%
N $, where the period $T=2\pi / \omega$. The frequency $\omega$ of the
driving field is tuned to match the splitting between the Zeeman doublet of
the discrete state, $\hbar \omega = E_2 - E_1$. Since the
driving is periodic, we can apply the Floquet technique \cite%
{Jiang2006,Reichl,KGS2012,KMSD2013,Demikhovskii2002} to obtain the stroboscopic picture of the system
state $\Psi(x,t)$ at discrete moments of time $t=NT$ \cite{Demikhovskii2002,Supplement}. 

When the wavefunction (\ref{psit}) is found, one
can calculate the stroboscopic evolution of various observables such as
\begin{equation}
x(t)=\int_{-\infty}^{\infty}\Psi^{\dagger}(x,t)x\Psi(x,t)dx, 
\end{equation}
for the position and 
\begin{equation}
\sigma_{z}(t)=\int_{-\infty}^{\infty}\Psi^{\dagger}(x,t)\sigma_{z}\Psi(x,t)dx, 
\end{equation}
for the spin component. 
We take the initial state $\Psi (x,0)$ as the ground state $E_{1}$ which has
the spin component $\sigma _{z}(0)$ close to one. When the first
zero of $\sigma _{z}(t)=0$ is achieved (as shown in Fig. \ref{figsztxt}(a))
during the stroboscopic evolution, we define this time as the half of the ``spin-flip'' time $T_{\rm{sf}}$.
For a few certain values of the system parameters and driving fields it is possible 
that the condition $\sigma _{z}(t)=0$ is never reached throughout the observation interval. 
In such a case we define the spin flip rate $1/T_{\rm{sf}}$ as equal to zero.

\subsection{Numerical results}

\subsubsection{System parameters}

We begin with setting the parameters for numerical calculations. We use a
donor state in InSb quantum wire. For InSb the value $m=0.0136m_{0}$ is
chosen for the electron effective mass where $m_{0}$ is the mass of a free
electron \cite{Saidi}. We assume $g=-50.6$ for 
the electron g-factor in InSb. The
amplitude of the Rashba SOC $\alpha $ in InSb can be tuned by the gate
voltage and can reach high values up to 100 meVnm \cite{Leontiadou,Wojcik}. 

For the shallow donor parameters we accept $U_{0}=1.5$ meV and $d=10$ nm,
as can be realized for a donor screened by a two-dimensional electron gas \cite{Ando}. 
As a result, a single discrete level $E_{0}=-0.072$ meV,
corresponding to the localization length $l$ close to 200 nm and frequency $|E_{0}|/\hbar\approx 2\pi \times 17.8$ GHz, 
is formed. All the states with positive energies belong to the continuum.
For the chosen parameters Eq. (\ref{bm}) gives the value $B_{\rm m}=25$ mT.  We consider three values of magnetic field: $%
B=0.25B_{\rm m},$ $B=0.5B_{\rm m},$ and $B=0.75B_{\rm m}$ and for each $B$ take two values
of SOC: $\alpha =6$ meVnm (for a relatively weak SOC 
with $l_{\rm so}={\hbar^{2}}/{m\alpha }\approx 10^{3}\mbox{ nm} > l$) 
and $\alpha =25$ meVnm (for a relatively strong SOC where $l_{\rm so} \approx l$).
The basis in Eq. (\ref{psit}), truncated at $l_{\rm max}=250 \ldots 500$, provides a 
good, $l_{\rm max}-$independent, convergence of all the numerical results in the range, 
where the continuum states are effectively involved in the evolution by the 
electric fields.

Condition of strong driving field amplitude $F_{s}\sim \left\vert{E_{0}}/{l}e\right\vert $ in Eq. (\ref{Vxt})
yields $\sim 5$ V/cm. Although this value is experimentally accessible, we will limit our calculations to
lower fields to avoid a nonlinear, strongly beyond the Fermi's golden rule, ionization process \cite{Delone}.  
Indeed, the semiclassical tunneling probability for a static field $F_{0}$ per time $\hbar/|E_{0}|$ can be estimated as
$\exp(-4F_{s}/3F_{0}),$ and one can introduce as a nominal reference parameter the tunneling ionization time
$\tau_{\rm ti}=\hbar/|E_{0}|\times \exp(4F_{s}/3F_{0}).$
Since the electric fields of our interest during the oscillation period are considerably 
weaker than $F_{s}$ the system is stable against the ionization.

\subsubsection{Time dependence of observables}

\begin{figure}[tbp]
\centering
\includegraphics*[width=65mm]{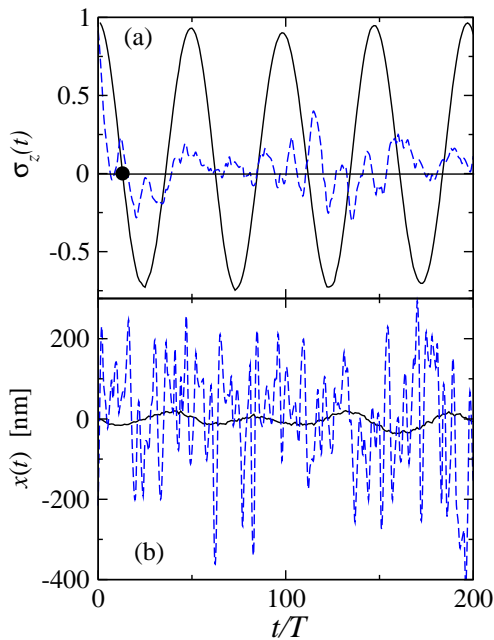}
\caption{Stroboscopic dynamics for mean values of (a) spin 
 component $\sigma_z(t)$ and (b) coordinate $x(t)$ for parameters $B=0.5B_{\rm m}$ and $\alpha=6$ meVnm. 
 The amplitude of the driving electric field  
$F_{0}=0.5$ V/cm (solid line) and $F_{0}=1.5$ V/cm (dashed line). Black circle in (a) corresponds to 
$T_{sf}/2$ for $F_{0}=0.5$ V/cm.
}
\label{figsztxt}
\end{figure}

To demonstrate a typical time dependence of the observables of interest, 
we show in Fig. \ref{figsztxt}  two examples of the evolution for mean values
 $\sigma _{z}(NT)$ and $x(NT)$ for driving fields with $F_{0}=0.5$ V/cm and 
$F_{0}=1.5$ V/cm, on the time interval $N<200$ of the 
driving field periods. Other parameters are $B=0.5B_{\rm m}$ and $%
\alpha =6$ meVnm giving the level splitting $E_2-E_1=0.035$ meV equal to $\approx 2\pi \times 8.65$ GHz of the driving frequency. It should be mentioned that this splitting includes the Zeeman coupling 
and the contribution due to the presence of the continuum states, which, at weak SOC 
is proportional to the square of the Rashba SOC strength (see Eq. (\ref{enshift})). In experiment, 
the driving frequency can be tuned smootly to match the exact level splitting which may differ 
from the simple Zeeman term. Note that this frequency, being a reference parameter, and independent of the driving field, does not include the dynamical Stark effect \cite{Stark}. 

It can be seen that at low driving
amplitude the dynamics of both spin and coordinate mean values shown in Fig.%
\ref{figsztxt} is rather regular, especially for the spin, where it
reminds the well-known picture for the Rabi resonance in a two-level system
with the frequency close to $e|x_{\lambda\lambda^{\prime}}|F_{0}/\hbar.$
As to the coordinate mean value, it demonstrates the combined drift and
oscillations which do not exceed the localization length, 
being of the order of $40-80$ nm. Thus, for low amplitudes
of driving the influence of delocalized continuum states of the nanowire is
relatively small both for spin and coordinate dynamics as it can be reduced mostly 
to the formation of nonzero matrix 
element of coordinate between localized spin-up and spin-down states.

The situation changes when the driving amplitude is increased to $F_{0}=1.5$ V/cm. 
It can be seen that both spin and coordinate
loose their simple evolution pattern visible at low driving field. They now
demonstrate more complicated dynamics with obviously many states
participating in it. This result clearly reflects the presence of continuum
states in the driven evolution, which become more significant when the
driving amplitude is increased. As to the numerical values in Fig. \ref{figsztxt}, 
one can see that the spin projection $\sigma_z(t)$ 
never approaches $-1$, i.e. no full spin flip is achieved. We
attribute this effect to the presence of many states with different spin
projections and with comparable weights, in the the total wavefunction (\ref{psit}). 
The coordinate mean value $x(t)$ at a strong driving
also demonstrates a complicated and irregularly oscillating behavior, with much greater
amplitude than for weak driving, approaching or even exceeding the localization length $l.$ 
This can be described as another evidence of significant
contribution of continuum states in the driven dynamics even for moderate
fields less then 3 V/cm.

\begin{figure}[tbp]
\centering
\includegraphics*[width=65mm]{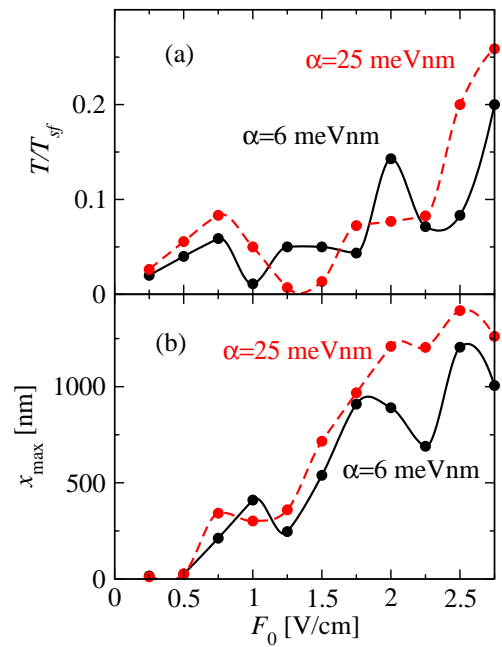}
\caption{Electric field amplitude dependence of (a) the
spin flip rate $1/T_{sf}$ and (b) the maximum displacement 
$x_{\max}$ for different parameters
of the spin-orbit coupling strength $\alpha$ and for $B=0.5 B_{%
\mathrm{m}}:$ (solid lines) $\alpha=6$ meVnm and (dashed lines) $%
\alpha=25$ meVnm. The lines serve only as the guides for the eye.}
\label{figb05}
\end{figure}

\begin{figure}[tbp]
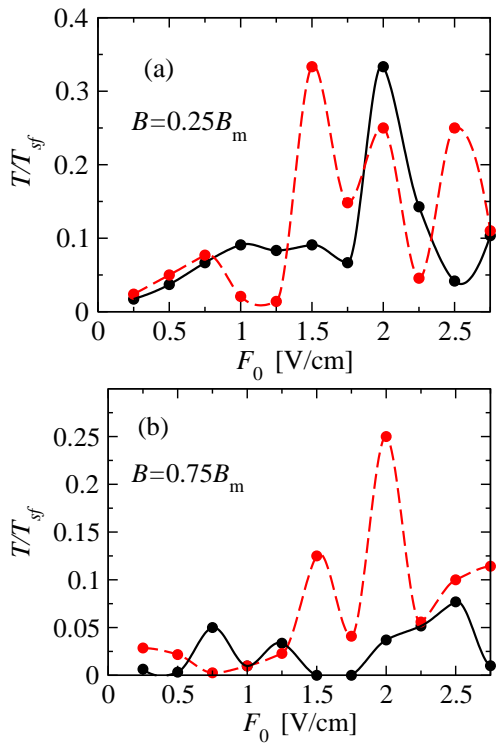

\centering
\includegraphics*[ width=65mm]{fig4a.eps}
\includegraphics*[ width=65mm]{fig4b.eps}
\caption{(a) Spin-flip rate for magnetic field $B=0.25 B_{\rm m}$. The parameters of the spin-orbit coupling
strength are $\alpha=6$ meVnm (solid line) and $\alpha=25$ meVnm (dashed line). 
(b) Same as in (a) but for higher magnetic field $B=0.75 B_{\rm m}$. The lines serve only as the guides for the eye.}
\label{figb025}
\end{figure}

\subsubsection{Dependence on the driving field amplitude}

It is of interest to track the amplitude characteristics of spin resonance
as functions of the driving field amplitude. First of all, we consider the
dependence of spin flip rate $1/T_{sf}$ and find it for several values of the
field amplitude $F_0$ as plotted in Fig.\ref{figb05}. Two values of
SOC coupling constant are taken for the magnetic field $B=0.5 B_{\rm m}$. One can see that for both
SOC amplitudes even at a moderate driving field amplitude $F_0 \approx 0.75$ V/cm,
the dependence of $1/T_{sf}$ on $F_0$ becomes strongly
nonlinear due to the significant role of continuum states in the driven
dynamics, taking the system out of a simple two-level approximation. Similar
behavior of the spin flip frequency in a multilevel system has been observed
in our previous studies of the spin resonance in a double quantum dot \cite{KGS2012}.
The complicated behavior of coupled spin and charge observables was also observed
in a multilevel system represented by a driven two-dimensional quantum billiard with SOC \cite{KMSD2013}.  
One can see in Fig. \ref{figb05} as well as in Fig. \ref{figb025} that
for certain values of the driving field the spin flip rate decreases to zero. The definition 
of such case has been discussed at the end of subsection IV.A, and it represents rare cases where the 
condition  $\sigma _{z}(t)=0$ is never reached during the observation time. 
The main cause of this effect cannot be assigned to the dynamical shift of the 
resonance condition \cite{Delone,Stark} since the significant 
presence of the continuum states with variable spin projections produces a much more complex coupled spin-position 
dynamics than that solely  caused by a time-dependent Stark effect. 
This is supported by an observation that the regions with zero $1/T_{\rm{sf}}$ occupy only few intermediate 
points for driving field amplitudes $F_{0}$ in 
Figs. \ref{figb05} and \ref{figb025}. One can see there that an increase in $F_{0}$ generates 
no results with vanishing  $1/T_{\rm{sf}}$ for $F_{0}$ close to the maximum values, 
so the observed feature is not definitely related to the dynamic energy level shift,
which grows with the increasing driving field.

Another characteristic of the driven evolution is the maximum achievable
displacement during the total observation time, $x_{\max},$ shown in Fig. \ref{figb05}(b) for
the same other parameters as in Fig. \ref{figb05}(a). One may expect that
the absolute value $x_{\max}$ grows up monotonically with
increasing the electric field strength $F_0$. 
However, strong SOC and the presence of the continuum provide sizable corrections to such 
a simple estimate, as it can be seen in Fig.\ref{figb05}. 
First of all, the dependence of $x_{\max}$ on $F_0$ demonstrates a nonlinear character 
from the fields of about $1$ V/cm for $\alpha=6$ meVnm and from $0.75$ V/cm for $\alpha=25$ meVnm. 
For the field amplitude exceeding $F_0 \approx 1.5$ V/cm, a saturation tendency is clearly visible, 
indicating that the electric field
efficiency in displacing the electron at large distance is
strongly reduced. 

One can be interested to look at the characteristics of the spin resonance
also for other values of the external magnetic field. Here we show the
results for the spin flip rate for the field $B=0.25 B_{\rm m}$ in Fig. \ref{figb025} (a) and
for $B=0.75 B_{\rm m}$ in Fig. \ref{figb025} (b).  As in Fig. \ref{figb05},
for each field we consider two values of the SOC amplitude $\alpha=6$ $%
\mathrm{meVnm}$ and $\alpha=25$ $\mathrm{meVnm}.$ 

By analyzing the results in Fig. \ref{figb05} and Fig. \ref{figb025},
one can conclude that both spin and coordinate dynamics have a
lot of common features for different parameters of the magnetic field and
SOC. The nonlinear dependence of $1/T_{\rm{sf}}$ and the 
$x_{\max}$ on the driving field amplitude is clearly visible. 
The maximum values of $1/T_{\rm{sf}}$ are comparable for all magnetic 
fields and for all SOC amplitudes. 
However, it can be seen from Fig. \ref{figb025}(b) that the increase in the 
magnetic field to $0.75 B_{\rm m}$ reduces the SOC-induced coupling between the states with different spin. 
Indeed, for this field, the values of $1/T_{\rm{sf}}$ are in general lower than those 
for $B=0.25 B_{\rm m}$ and $B=0.5 B_{\rm m}$. 
This finding is especially clear for the low SOC strength $\alpha=6$ meVnm. As one can see in all the Figures,
the spin-flip time $T_{\rm{sf}}$ usually does not exceed the nominal tunneling ionization time $\tau_{\rm ti},$
confirming that the time scale of the spin flip process is usually shorter than the time scale of
the tunneling ionization.\cite{Supplement}

The two values of Rashba coupling considered here represent its actual operating range 
in prospective systems based on InSb nanowires. Indeed, neither very small nor very large SOC is desirable. 
If the SOC is too weak, the coupling between spin-resolved discrete states via the continuum becomes ineffective. 
For a very strong coupling one needs large amplitudes  $F_{0}$ to 
flip the spin and to make the qubit actually working since highly mixed spin states 
are involved and produced by the driving. As a result, a certain range of the SOC strength 
is needed as represented by these two realizations.

Another key parameter is the 
driving field amplitude $F_{0},$ limited here by a moderate value of 2.75 V/cm. 
The first reason for keeping it not very strong is avoiding the nonlinear ionization \cite{Delone}.
The second reason is the absence of an effective increase 
for the spin flip rate when making the driving amplitude too strong, as it can be seen 
in Figs. \ref{figb05} and \ref{figb025}. When the driving is too intense, the 
continuum states with different spins begin to play more significant 
role by creating a hardly distinguishable mixture of the two discrete states with 
the continuum. In such a mixture the spin components together with the wavefunction 
spread may become poorly defined, prohibiting the application of such a regime for a 
practically operating qubit. Besides, strong electric fields are in general not desirable for 
micron- and submicron-sized electronic devices. This is why we restrict ourselves to low and medium driving 
fields, sufficient for providing an effective operation for the qubit.

Here one additional comment might be of interest. Although the behavior of the driven spin dynamics in the regime 
of strong spin-orbit coupling is practically unpredictable, some conclusions can be made
from the extension of the perturbation theory analysis. In this case, the matrix element of the coordinate
$x_{\lambda\lambda^{\prime}}$ is of the order of the localization length $l$. Therefore, the spin-flip Rabi
frequency of the order of $eF_{0}l/\hbar$ sets the upper limit on the spin-flip rate for a relatively weak 
external field, which is confirmed by our results in Fig. \ref{figsztxt} for lower amplitude $F_{0}$. For a strong 
field and a relatively weak spin-orbit coupling, we can conclude from the 
matrix element on Eq. (\ref{xintegral}) that the states providing the maximal contribution 
to (\ref{xintegral}) and to the spin dynamics have momenta of the 
order of $\hbar/l$. Therefore, they are described by the precession rate of the order of $\alpha/\hbar l$ 
that may help to estimate 
the actual spin-flip rate. However, it can be seen in Figs. \ref{figb05} and \ref{figb025} 
that an increase in the SOC amplitude alone does not lead to the strictly proportional growth of 
the spin flip rate since the dynamics has a 
complex nature with many states participating in it.

We note that in all the regimes considered here, $T_{\rm{sf}}$ is of the order of or less than $10 T$.
Taking into account that $T$ is of the order of 0.1 ns, we find that $T_{\rm{sf}}$ is of the order of
a nanosecond. Since the spin relaxation time in week magnetic fields exceeds this time by
orders of magnitude \cite{Khaetsky,Amasha}, at the time scale of $N\sim 100$, we still have a 
coherent qubit manipulation in the EDSR regime. Note that the effect of the noise in the electric 
field always present in the semiconductor system of interest, will not change this result since 
the spectrum of the noise is not peaked in the frequency range corresponding to
the EDSR in the magnetic fields considered in this paper.\cite{Li2018}

\section{Conclusions}

We studied the electric dipole spin resonance for a nanowire-based donor states coupled to the continuum, the latter playing a critical role in the dynamics. The continuum leads to 
a strongly nonlinear dependence of the evolution of spin and position on the electric field.
The observed characteristics of both spin and position 
dynamics, having much in common, for different values of magnetic field
and spin-orbit coupling, can be of interest for designing novel types of spin and charge
qubits when the confining potentials are shallow, and the discrete states strongly
interact with the continuum during the qubit operation. For this reason, 
these effects should be taken into account for possible
applications of materials with strong spin-orbit coupling such as InSb for fabricating the
qubit-processing structures.

\section*{Acknowledgements}

D.V.K. and E.A.L. are supported by the State Assignment of the Ministry of
Education and Science RF (project No. 3.3026.2017/PCh). 
E.A.L. was supported by the grant of the President of the 
Russian Federation for young scientists MK-6679.2018.2.
E.Y.S. acknowledges support by the 
Spanish Ministry of Economy, Industry and Competitiveness (MINECO) 
and the European Regional Development Fund FEDER through Grant No. FIS2015-67161-P (MINECO/FEDER, UE), 
and the Basque Government through Grant No. IT986-16.


\begin{thebibliography}{99}
\bibitem{Lparameter}One can estimate the number of continuum states 
${\cal N}$ occupying the typical energy interval of the order of $|E_{0}|$ as
${\cal N}\sim L/l\gg 1.$ This inequality assures a large density of
the extended states and represent a validity approximation for the continuum.
In our numerical studies we took this ${\cal N}\approx 10^{2}$
which gave us a reliable approach for the continuum states.
\bibitem{Stark} In the absence of the SOC, in a weak static electric field $F$ both spin-split localized
states experience a Stark shift $E_{0}(F/F_{s})^{2},$ as can be found by perturbation theory using
Eq. (\ref{xintegral}). Their second-order in $\alpha$ Stark-originated splitting 
is then of the order of $E_{0}(F/F_{s})^{2}\times(l/l_{\rm so})^2.$ As a result, for the set of  
magnetic fields of our interest, the direct 
effect of the relative Stark shift compared to the effective Zeeman splitting  
becomes important if both the driving field and the spin-orbit coupling 
are relatively strong.
\end{thebibliography}
\end{document}